\input harvmac


%
\def\eql{~=~}
\def\seql{\! = \!}

\def\coeff#1#2{\relax{\textstyle {#1 \over #2}}\displaystyle}
\def\half{{1 \over 2}}

 \def\cG{{\cal G}}
 
\def\cJ{{\cal J}} 
 \def\cM{{\cal M}}
\def\cN{{\cal N}} \def\cO{{\cal O}}
\def\cP{{\cal P}}

\def\cV{{\cal V}} \def\cW{{\cal W}}
\def\cX{{\cal X}} 

\def\bfone{\relax{\rm 1\kern-.35em 1}}
\def\inbar{\vrule height1.55ex width.4pt depth0pt}
\def\IC{\relax\,\hbox{$\inbar\kern-.3em{\rm C}$}}
\def\ID{\relax{\rm I\kern-.18em D}}
\def\IF{\relax{\rm I\kern-.18em F}}
\def\IH{\relax{\rm I\kern-.18em H}}
\def\II{\relax{\rm I\kern-.17em I}}
\def\IN{\relax{\rm I\kern-.18em N}}
\def\IP{\relax{\rm I\kern-.18em P}}
\def\IQ{\relax\,\hbox{$\inbar\kern-.3em{\rm Q}$}}
\def\IR{\relax{\rm I\kern-.18em R}}
\def\us#1{\underline{#1}}
\font\cmss=cmss10 \font\cmsss=cmss10 at 7pt
\def\ZZ{\relax\ifmmode\mathchoice
{\hbox{\cmss Z\kern-.4em Z}} {\hbox{\cmss Z\kern-.4em Z}}
{\lower.9pt\hbox{\cmsss Z\kern-.4em Z}}
{\lower1.2pt\hbox{\cmsss Z\kern-.4em Z}}\else{\cmss Z\kern-.4em Z}\fi}

\def\nihil#1{{\it #1}}
\def\eprt#1{{\tt #1}}
\def\nup#1({Nucl.\ Phys.\ $\us {B#1}$\ (}
\def\plt#1({Phys.\ Lett.\ $\us  {#1B}$\ (}
\def\cmp#1({Comm.\ Math.\ Phys.\ $\us  {#1}$\ (}
\def\prp#1({Phys.\ Rep.\ $\us  {#1}$\ (}
\def\prl#1({Phys.\ Rev.\ Lett.\ $\us  {#1}$\ (}
\def\prv#1({Phys.\ Rev.\ $\us  {#1}$\ (}
\def\mpl#1({Mod.\ Phys.\ Let.\ $\us  {A#1}$\ (}
\def\ijmp#1({Int.\ J.\ Mod.\ Phys.\ $\us{A#1}$\ (}
\def\jag#1({Jour.\ Alg.\ Geom.\ $\us {#1}$\ (}

%


%
%
\lref\KPW{A.\ Khavaev, K.\ Pilch and N.P.\ Warner, \nihil{New
Vacua of Gauged  ${\cal N}=8$ Supergravity in Five Dimensions},
\eprt{hep-th/9812035}.}
\lref\DistZam{J. Distler and F. Zamora, \nihil{Non-supersymmetric
conformal field theories from stable anti-de Sitter  spaces},
Adv. Theor. Math. Phys. {\bf 2} (1999) 1405, \eprt{hep-th/9810206};
\nihil{Chiral symmetry breaking in the AdS/CFT correspondence},
\eprt{hep-th/9911040}.}
\lref\GPPZold{L. Girardello, M. Petrini, M. Porrati and A.
Zaffaroni \nihil{Novel local CFT and exact results on
perturbations of N = 4 super  Yang-Mills from AdS dynamics},
JHEP {\bf 12} (1998) 022, \eprt{hep-th/9810126}.}
\lref\FGPWa{D.~Z. Freedman, S.~S. Gubser, K.~Pilch, and N.~P. Warner,
\nihil{Renormalization Group Flows from Holography---Supersymmetry
and a c-Theorem,} CERN-TH-99-86, \eprt{hep-th/9904017} }
\lref\FGPWb{D.~Z. Freedman, S.~S. Gubser, K.~Pilch, and N.~P. Warner,
{\it Continuous Distribution of D3-branes and Gauged Supergravity,}
\eprt{hep-th/9906194}. }
\lref\GPPZ{L. Girardello, M. Petrini, M. Porrati and A.
Zaffaroni \nihil{The supergravity dual of N = 1 super Yang-Mills theory
,}
Nucl. Phys. {\bf B569} (2000) 451, \eprt{hep-th/9909047}.}
\lref\Beh{K.~Behrndt,
\nihil{Domain walls of D = 5 supergravity and fixpoints of N = 1 super
Yang-Mills}, \eprt{hep-th/9907070}.}
\lref\SGsing{S.~Gubser, \nihil{Curvature Singularities:  The Good,
The Bad, and the Naked,} PUPT-1916,
\eprt{hep-th/0002160}.}
\lref\NWrev{N.P.\ Warner, \nihil{Renormalization Group Flows from
Five-dimensional Supergravity,} talk presented at Strings `99,
Potsdam, Germany, 19-25 Jul 1999, Class.~Quant.~ Grav. {\bf 17}
(2000),  1287;  \eprt{hep-th/9911240}.}
\lref\PetZaf{M.~Petrini and A.~Zaffaroni,
\nihil{The holographic RG flow to conformal and non-conformal theory},
\eprt{hep-th/0002172}.}
\lref\NEMP{N.~Evans and M.~Petrini, \nihil{AdS
RG Flow and the Super-Yang-Mills Cascade,}  SHEP-00-05,
IMPERIAL-TP-99-00-28,  \eprt{hep-th/0006048}.}
\lref\GRW{M.\ G\"unaydin, L.J.\ Romans and N.P.\ Warner,
\nihil{Gauged $N=8$ Supergravity in Five Dimensions,}
Phys.~Lett.~{\bf 154B} (1985) 268; \nihil{Compact and Non-Compact
Gauged Supergravity Theories in Five Dimensions,}
\nup{272} (1986) 598.}
\lref\PPvN{M.~Pernici, K.~Pilch and P. van Nieuwenhuizen,
\nihil{Gauged $N=8$, $D = 5$ Supergravity,} \nup{259} (1985) 460.}
\lref\RLMS{R.~G. Leigh and M.~J. Strassler, \nihil{Exactly Marginal
Operators and Duality in Four-Dimensional $N=1$ Supersymmetric 
Gauge Theory,} \nup{447} (1995) 95; \eprt{hep-th/9503121}.}
\lref\KPNW{K.~Pilch and N.P.~Warner, \nihil{$\cN=2$ Supersymmetric 
RG Flows and the IIB  Dilaton,} CITUSC/00-18, USC-00/02; 
\eprt{hep-th/0004063}.}
%

\Title{ \vbox{ \hbox{CITUSC/00-051} \hbox{USC-00/04} \hbox{\tt
hep-th/0009159} }} {\vbox{\vskip -1.0cm
\centerline{\hbox
{A Class of ${\cal N} \!=\! 1$ Supersymmetric RG Flows}}
\vskip 8 pt
\centerline{
\hbox{from Five-dimensional ${\cal N}  \!=\!  8$ Supergravity}}}}
\vskip -.3cm
\centerline{Alexei Khavaev and Nicholas P.\ Warner }
\medskip
\centerline{{\it Department of Physics and Astronomy}}
\centerline{{\it and}}
\centerline{{\it CIT-USC Center for
Theoretical Physics}}
\centerline{{\it University of Southern California}}
\centerline{{\it Los Angeles, CA 90089-0484, USA}}

\bigskip
\bigskip
We consider the holographic dual of
a general class of ${\cal N} \seql 1^*$ flows
in which all three chiral multiplets have independent
masses, and in which the corresponding Yang-Mills scalars can
develop particular supersymmetry-preserving vevs.
We also allow the gaugino to develop a vev.  This
leads to a six parameter subspace of the supergravity
scalar action, and we show that this is a consistent
truncation, and obtain a superpotential that governs
the ${\cal N}=1^*$ flows on this subspace.  We analyse
some of the structure of the superpotential,  and check
consistency with the asymptotic behaviour near the
UV fixed point.  We show that the dimensions of the
six couplings obey a sum rule all along
the ${\cal N} \seql 1^*$ flows.   We also show how our
superpotential describes part of the Coulomb branch of the
non-trivial ${\cal N}=1$ fixed point theory.

\vskip .3in
\Date{\sl {September, 2000}}

\parskip=4pt plus 15pt minus 1pt
\baselineskip=15pt plus 2pt minus 1pt

\newsec{Introduction}

The study of holographic RG flows has become one
of the more enduring spin-offs of the AdS/CFT correspondence.
In particular the flows of  $\cN=4$ supersymmetric
Yang-Mills theories under ``bilinear perturbations\foot{
By this we mean either the introduction of mass terms
for fundamental fields, or the turning on of vevs for
bilinear operators.}''  has been extensively studied
\refs{\KPW\DistZam \GPPZold\FGPWa\FGPWb\GPPZ\Beh\SGsing
\NWrev\PetZaf-\NEMP} using gauged $\cN=8$ supergravity in five
dimensions \refs{\GRW,\PPvN}.
The most extensive studies have been made for supersymmetric
flows because these are generically simpler, more
controllable,  and sometimes enable direct comparisons
of the supergravity and field theory limits.
Thus far the flows have usually been the simpler ones
with a higher degree of global symmetry,
\refs{\DistZam\GPPZold\FGPWa\FGPWb-\GPPZ} though a relatively 
recent paper \NEMP\ has pushed this restriction back even 
further.

In this letter we will give the supergravity description
of an even more general class of  flows of
$\cN \seql 4$ supersymmetric Yang-Mills: $\cN\seql 1$
supersymmetric flows with six independent parameters, which
may be interpreted as three independent masses of the chiral
multiplets, two independent vevs of the scalar fields in these
chiral mutiplets, and the vev of the gaugino condensate.
We work with gauged $\cN\seql 8$ supergravity in five dimensions,
and obtain a superpotential describing these flows via
steepest descent.  Indeed we find four different superpotentials
that are trivially related by a $\ZZ_4$ symmetry:
Each of the four superpotentials merely corresponds to 
a different fermion being identified as the gaugino.

While we have the supergravity description of flows with all
three chiral multiplet masses running independently, this
is not the most general class of $\cN=1$ supersymmetric flow
under ``bilinear perturbations'': It is still possible to flow
the scalar and fermion  bilinear vevs in more exotic ways
than those considered here.  On the other hand, we show that
the six parameter truncation that we consider here is
a consistent truncation of the supergravity model, and
thus the flows involve a closed family of  bilinear operators.
We anticipate that our results will prove valuable in further
probing holographic renormalization group flows.  In this
letter we will content ourselves with exhibiting the
superpotential, showing that there are no new critical points,
obtaining a sum-rule for the anomalous dimensions of fermion
bilinears, and exhibiting the Coulomb branch of the $\cN \seql 1$
supersymmetric Leigh-Strassler point \RLMS.

\newsec{The scalar manifold}

\subsec{A non-compact Cartan sub-algebra}

The forty-two scalars of  $\cN \seql 8$ supergravity
parametrize non-compact coset
space $E_{6(6)}/USp(8)$, and can be described by a $27 \times 27$
matrix, ${\cV_{AB}}^{ab}$.  Working in the $SL(6,\IR)\times SL(2,\IR)$
basis, an element of the $E_{6(6)}$ Lie algebra can be written in the
block form as \GRW\
\eqn\Xdef{ \cX \eql \left( \matrix{
-4 \, \Lambda^{[M}{}_{[I}\delta^{N]}{}_{J]} &
\sqrt{2} \, \Sigma_{IJP\beta} \cr
\sqrt{2} \, \Sigma^{MNK\alpha} &
\Lambda^K_P\, \delta^{\alpha}{}_{\beta} \, + \,
\Lambda^{\alpha}_{\beta}\, \delta^{K}{}_{P} }\right) \ ,}
where $\Lambda^K_P$, $\Lambda^\alpha_\beta$  represent
elements of the $SL(6,\IR)\times SL(2,\IR)$ Lie algebra and
$\Sigma_{IJP\beta}$ transforms in the ${\bf (20,2)}$ of
$SL(6,\IR)\times SL(2,\IR)$.    The scalar fields to which
we wish to restrict are most easily characterized as
follows.  Introduce Cartesian coordinates $x^I, I =1,\dots, 6$,
and $y^\alpha, \alpha =1, 2$,
and define the differential form $\Sigma$ by:
\eqn\sigdefn{\Sigma ~=~ {1 \over 12}\, \Sigma_{IJK\alpha} \,
dx^I \,\wedge \, dx^J \,\wedge \, dx^K \,\wedge \, dy^\alpha \ .}
Now introduce complex coordinates $z_1 = x^1 + i x^2$,
$z_2 = x^3 - i x^4$, $z_3 = x^5 - i x^6$, and $z_4 = y^1 + i y^2$.
Four of the generators we seek can be defined by setting:
\eqn\siglincomb{ \Sigma ~=~ k\, \sum_{i=1}^4 \, \varphi_i \
(~\Upsilon_i ~+~ \overline{\Upsilon_i}~) \, ,}
where $k$ is a normalization constant and
\eqn\upsdefn{\eqalign{ \Upsilon_1 ~\equiv~ & -dz_1 \wedge dz_2
\wedge dz_3 \wedge dz_4 \, ,\qquad
\Upsilon_2 ~\equiv~  -d\overline{z}_1 \wedge d\overline{z}_2
\wedge dz_3 \wedge dz_4 \cr
\Upsilon_3 ~\equiv~ & -d\overline{z}_1 \wedge dz_2 \wedge
d\overline{z}_3 \wedge dz_4 \, , \qquad
\Upsilon_4 ~\equiv~  -dz_1 \wedge d\overline{z}_2 \wedge
d\overline{z}_3  \wedge dz_4 \,.}}
The remaining two generators are given by taking the
$SL(6,\IR)$ Lie algebra element to be of the form:
\eqn\Lamdefn{\Lambda ~\equiv~ {\rm diag} (\alpha+\beta,
\alpha+ \beta, \alpha -\beta, \alpha -\beta,
-2\,\alpha,-2\, \alpha,) \,.}
One can easily verify that the foreging six generators
constitute a (non-compact) Cartan sub-algebra of
$E_{6(6)}$, and indeed one can show that they
constitute the diagonal elements of another
$SL(6,\IR)\times SL(2,\IR)$ sub-algebra (distinct
from the one described above).

The kinetic term for these scalars is:
\eqn\scalkin{-3\, (\del \alpha)^2 ~-~ (\del \beta)^2 ~-~
\coeff{1}{2}\, \sum_{j=1}^4 \, (\del \varphi_j)^2  \,,}
where the normalization constant $k$ in \siglincomb\
has been chosen so as to canonically normalize the
kinetic terms of the $\varphi_j$.

In terms of the theory on the brane, the parameters
$\varphi_i$ are dual to the operators:
$$
 \Tr (\lambda_i \lambda_i) \, + \, h.c.  \,,
$$
while $\alpha$ and $\beta$ are, respectively, dual to
\eqn\scalops{\eqalign{
\cO_1 ~\equiv~ & \Tr (X_1^2 + X_2^2 +X_3^2 + X_4^2 - 2\,
X_5^2 -2\, X_6^2)  \quad {\rm and} \cr
\quad \cO_2 ~\equiv~ &\Tr (X_1^1 + X_2^2 - X_3^2 - X_4^2) \,.}}
It should also be remembered that the operator:
\eqn\Ozero{\cO_0 ~\equiv~ \Tr \Big(\sum_{i=1}^6 X_i^2
\Big) \,,}
has no supergravity dual in the gauged $\cN=8$ supergravity theory,
but that the field theory on the brane always adds an appropriate
amount of $\cO_0$ to the operators $\cO_j, j=1,2$ so as to
preserve supersymmetry and positivity.

\subsec{Consistency of the truncation}

The simplest way to establish the consistency
of a truncation is to exhibit a symmetry (continuous or
discrete) of the action for which the truncated sector consists
of precisely all the singlet fields under that symmetry.
It then follows from Schur's lemma that all variations
of the action must be at least quadratic in
non-singlet fields, and hence it is consistent
with the equations of motion to set all non-singlets
fields to zero.  We now exhibit a symmetry that
reduces the forty-two scalars to the six described
above.

Consider the following two $\ZZ_2$ generators in
$SO(6)$:
\eqn\Ztwos{
diag(-1, -1, -1, -1, 1, 1) \quad \hbox{ and } \quad
diag(1, 1, -1, -1, -1, -1)  \,.}
This $\ZZ_2 \times \ZZ_2$ can be viewed as simultaneously 
negating pairs of the complex coordinates $z_i$.  As a result,
demanding invariance under these symmetries of the
potential means that $\Lambda^I_J$ must be block diagonal
with three $2 \times 2$ blocks.  Similarly, it requires
that $\Sigma$ contain one of each pair $\{dz_i,d\bar z_i\},
i=1,\dots,3$.  Now introduce the matrices:
$$
\cJ_1 \eql \left( \matrix{
   0 & 1 & & &    &        \cr
  -1 & 0 & & &    &        \cr
     &   & 0 & -1 & &      \cr
     &   & 1 &  0 & &      \cr
     &   &   &    & 0 & -1 \cr
     &   &   &    & 1 &  0 \cr } \right) \,, \qquad
\qquad
\cJ_2 \eql \left( \matrix{ 0 & 1 \cr
  -1 & 0 \cr } \right) \,,
$$
considered as elements of $SO(6) \times SO(2) \subset
SL(6,\IR) \times SL(2,\IR)$.
The potential is invariant under the action of both of these
matrices.  Consider the symmetry group, $\cG$, generated
by the  $\ZZ_2 \times \ZZ_2$ of \Ztwos, along with the
combined action $\cJ_1 \cJ_2$.  The latter generates a $\ZZ_4$,
and invariance requires that the $2 \times 2$ blocks in
$\Lambda^I_J$ are in fact independent multiples of
the $2 \times 2$ identity matrix.  Remembering that
$\Lambda^I_J$ is traceless, we have thus reduced it to
\Lamdefn.  The combined action of $\cJ_1 \cJ_2$ rotates
$z_a \to -i z_a, a=1,\dots,4$, and invariance under
this (and the other $\ZZ_2$'s) reduces $\Sigma$ to
a linear combination of the $\Upsilon_i$ and their
complex conjugates \upsdefn.   We now observe that
$\Lambda$ in \Lamdefn\ commutes with an $SO(2)\times SO(2)
\times SO(2) \subset SO(6)$, and that these $U(1)$'s, along
with the $U(1) \subset SL(2,\IR)$ can be used to
reduce to the real linear combination in \siglincomb.
Putting it another way, the discrete symmetry
reduces the the $E_{6(6)}/Usp(8)$ coset to
\eqn\fnlcoset{\cM \eql \bigg({SU(1,1) \over U(1)}\bigg)^4 ~
\times~ SO(1,1) ~\times~  SO(1,1) \,,}
where the denominator $U(1)^4 \subset SO(6) \times SO(2)$,
the numerator $SU(1,1)$'s are paramet\-rized by the
$\Upsilon_a$, and the  $SO(1,1)$'s are parametrized by
$\alpha$ and $\beta$.  The $U(1)$'s can then be used
to gauge fix the complex parameter in each $SU(1,1)$
to be real.

\newsec{The supergravity superpotential and supersymmetric flows}

\subsec{The superpotential}

The superpotential is generically extracted
as an eigenvalue of the tensor, $W_{ab}$, of the
five-dimensional $\cN=8$ theory.  For the scalar
manifold that we consider here, the eigenvectors
are all constant vectors, and indeed are relatively
simple:
\eqn\evecs{\eqalign{\vec v_1 \eql & (-1, 0, 1, 0, 0, 1, 0, 1)
\,, \qquad \vec v_2 \eql (0, 1, 0, 1, 1, 0, -1, 0) \,, \cr
 \vec v_3 \eql & ( 1, 0, 1, 0, 0, -1, 0, 1) \,, \qquad
 \vec v_4 \eql (0, 1, 0, -1, 1, 0, 1, 0) \,, \cr
\vec v_5 \eql & (1, 0, -1, 0, 0, 1, 0, 1) \,, \qquad \vec v_6
\eql  (0, 1, 0, 1, -1, 0, 1, 0) \,, \cr
 \vec v_7 \eql & (1, 0, 1, 0, 0, 1, 0, -1) \,, \qquad
\vec v_8 \eql (0, -1, 0, 1, 1, 0, 1, 0) \,.}}
There are four distinct eigenvalues corresponding to the
spaces spanned by $\vec v_{2a-1}, \vec v_{2a }$, for
$a=1,\dots,4$.  The first eigenvalue is:
\eqn\superpot{\eqalign{\cW  \eql {1 \over 4}\, \Big[\,& 
\big( \rho^{-4} - \rho^2 \,(\nu^2 + \nu^{-2})\big) \,
\cosh(2\, \varphi_1) ~+~ \big(- \rho^{-4} + \rho^2 \,
(- \nu^2 + \nu^{-2})\big) \, \cosh(2\, \varphi_2) ~+~ \cr
&  \big(- \rho^{-4} + \rho^2 \,(\nu^2 - \nu^{-2})\big) \,
\cosh(2\, \varphi_3) ~-~ \big(\rho^{-4} + \rho^2 \,(\nu^2 +
\nu^{-2})\big) \,\cosh(2\, \varphi_4) \,\Big]\,,} }
where $\rho \equiv e^\alpha$ and $\nu \equiv e^\beta$.
The remaining three eigenvalues are obtained from this by
doing the pairwise permutations:
\eqn\pairswitch{\varphi_1 \leftrightarrow \varphi_4 \,,
\varphi_2 \leftrightarrow  \varphi_3\, ; \qquad
\varphi_1 \leftrightarrow \varphi_3 \,,\ \varphi_2
\leftrightarrow  \varphi_4\,; \quad {\rm and} \quad \varphi_1
\leftrightarrow  \varphi_2 \,, \ \varphi_3 \leftrightarrow
\varphi_4\,.}

As one would hope, $\cW$ does indeed provide a superpotential
in that the supergravity potential, $\cP$, is given by:
\eqn\spiden{
\cP \eql
{1 \over 8} \,\sum_{i=1}^4 \bigg({\partial\cW \over 
\partial\varphi_i} \bigg)^2 ~+~ {1 \over 48}\, 
\bigg({\partial\cW \over \partial \alpha} \bigg)^2 ~+~  
{1 \over 16}\, \bigg({\partial\cW \over \partial\beta} 
\bigg)^2 ~-~  {1 \over 3}\, {\cW}^2\,. }
Indeed, because the actions \pairswitch\ interchange
the eigenvalues, it follows that \spiden\ yields the
supergravity potential, $\cP$, if $\cW$ is chosen
to be any of the eigenvalues of $W_{ab}$.

One should also note that $\cW$ is {\it invariant, up
to a sign,} under the
permutation group $S_3$.  These permutations are generated
by the transformations:
\eqn\permgens{\eqalign{ p_1 ~:~ &\quad \varphi_1 \leftrightarrow
\varphi_2 \,, \quad \varphi_3 \to \varphi_3 \,,  \quad
\alpha \to -\coeff{1}{2}(\alpha+\beta) \,, \quad \beta \to
\coeff{1}{2}(\beta - 3\,\alpha)  \cr
p_2 ~:~ &  \quad \varphi_2 \leftrightarrow
\varphi_3 \,, \quad \varphi_1 \to \varphi_1 \,, \quad
\alpha \to \alpha\,, \quad \beta \to -\beta\,,}}
and these act on $\cW$ according to: $p_1:\cW \to -\cW$ and
$p_2:\cW \to \cW$.

In terms of the physics on the brane, the choice of
the eigenvalue of $W_{ab}$ corresponds to the choice
of which of the four fermions will be the gaugino.  The
permutation symmetry generated by \permgens\ represents
the permutations of the three chiral multiplets.  Thus
the physical content of all four possible superpotentials
is the same, and we therefore stay with the choice \superpot.

\subsec{Supersymmetric flows}

As is common when considering supergravity descriptions
of RG flows, we will take the five-dimensional metric to have
the form
\eqn\RGFmetric{
ds^2_{1,4} = e^{2 A(r)} \eta_{\mu\nu} dx^\mu dx^\nu - dr^2 \,.}
For a supersymmetric flow one requires that the variations
of the spin-${3 \over 2}$ and spin-${1 \over 2}$ fields to vanish.
The former leads to the condition:
\eqn\cosmic{{d A \over d r}\eql  - {g \over 3}\,W \ .}
The variations of the spin-${1 \over 2}$ fields lead to the
equations:
\eqn\floweqs{\eqalign{ {d \alpha \over d r} \eql  &{g \over 12}\,
{\del W \over \del \alpha}  \ , \qquad {d \beta \over d r}
\eql  {g \over 4}\,  {\del W \over \del \beta} \ ,\cr
{d \varphi_j \over d r} \eql  &{g \over 2}\, {\del W \over \del
\varphi_j } \ , \qquad  j=1,\dots,4 \ .}}
For later convenience we introduce the canonically normalized
scalars $\varphi_5 \equiv \sqrt{6}\, \alpha$,
$\varphi_6 \equiv \sqrt{2}\, \beta$.

It is straightforward, but rather tedious to verify directly that
these equations imply that all the spin-${1 \over 2}$ variations
vanish.  We have indeed confirmed this using
{\it  Mathematica}$^{TM}$,
however there is a rather general analytic argument that
establishes the same result.

Suppose that $\xi^a$ is an eigenvector of $W_{ab}$, with
eigenvalue $\cW$.  Let $\zeta^a \equiv \Omega^{ab} (\xi^b)^*$,
where $\Omega$ is the symplectic form defined in \GRW.
{}From the symplectic reality condition satisfied by
$W_{ab}$ it follows that $\zeta^a$ must be an eigenvector
with eigenvalue $\cW^*$.  For simplicity, assume that
$\xi, \zeta$ and $\cW$ are real (as they are here).

To prove that the spin-${1 \over 2}$ variations vanish we must
show that the following expression vanishes:
\eqn\spinhalf{P_{0\,abcd} \, \xi^d ~\pm~ \half\,g \,A_{dabc}\,
\zeta^d \ ,}
for one uniform choice of the sign. The quantities in this
equation are defined in \GRW.  We consider the modulus
squared of \spinhalf.  Recalling that the
canonically normalized kinetic term is ${1 \over 24}
|P_{\mu \,abcd}|^2$, it is relatively easy to show that:
\eqn\Ppart{ {1 \over 24}\,P_{0\,abcd} {P_0}^{abce} \,
\xi^d \, \xi^e \eql
{1 \over 16} \, \sum_{j=1}^6 \, (\del \varphi_j)^2 \ .}
The factor of ${1 \over 8}$ compared to the usual kinetic
term arises from the fact that we are not taking
the trace over the last index, but we simply
contract with $\xi$.

The cross-terms between $P_{0\,abcd}$ and $A_{dabc}$
can be simplified using the identity (3.21) of \GRW:
\eqn\cterms{D_\mu \, W_{ab} ~=~ {2 \over 3} \,
{P_{\mu \, (a}}^{cde} \, A_{b)cde}  \ .}
Finally, contracting the following identity ((3.31) of \GRW):
\eqn\quadid{{1 \over 24}\, W^{ac} \, W_{cb} ~-~
{1 \over 96}\, A^{acde} \, A_{bcde}
\eql -{1 \over 8 g^2}\, \cP \, \delta^a_b \ ,}
with $\xi^b$, and using equation \spiden, it follows
immediately that:
\eqn\Asquare{ A^{acde} \, A_{bcde} \, \xi^b \eql
{3 \over 2} \, \bigg(\,\sum_{i=1}^6 \, {\del \cW \over
\del \varphi_i} \,\bigg) \, \xi^a\ .}
Putting together \Ppart, \cterms\ and \Asquare, we arrive
at:
\eqn\susycond{ \Big|\,P_{0\,abcd} \, \xi^d ~\pm~ \half\,g \,
A_{dabc}\, \zeta^d \,\Big|^2 ~=~ {3 \over 2}\, \big|\xi^a\big|^2 \,
\sum_{i=1}^6 \, \bigg|\,{d \varphi_i \over d r} ~\pm~
{1 \over 2}\, g\, {\del \cW \over \del \varphi_i} \,\bigg|^2 \, .}
It therefore follows that the spin-${1 \over 2}$ variations
vanish if and only if the steepest descent equations in
\floweqs\ are satisfied.

\newsec{Properties of the Superpotential}

\subsec{Behaviour near the maximally supersymmetric phase}

In the maximally supersymmetric phase all the supergravity
scalars vanish.  In the neighbourhood of this point we find:
\eqn\asympW{\eqalign{{\cal W} ~\sim~ & -{3 \over 2} ~-~
{1 \over 2}\,(\varphi_1^2  + \varphi_2^2 +\varphi_3^2 + 3\,
\varphi_4^2 + 2 \, (\varphi_5^2 + \varphi_6^2)) \cr & ~-~
\sqrt{{2 \over 3}}\  \varphi_5\, (2\, \varphi_1^2 -\varphi_2^2 -
\varphi_3^2)
~-~ \sqrt{2} \, \varphi_6\, (\varphi_2^2 -\varphi_3^2) \ ,}}
where, one should recall, that  $\varphi_5 \equiv \sqrt{6}\, \alpha$,
$\varphi_6 \equiv \sqrt{2}\, \beta$ are the canonically normalized
supergravity scalars.

The quadratic terms imply that perturbations in $\varphi_1, \varphi_2$
and $\varphi_3$ are non-normalizable in $AdS_5$,
and so represent non-trivial masses for the corresponding fermion
fields.  The quadratic terms in $\varphi_{j}, j=5,6$ give rise to
normalizable $AdS_5$ modes, however, as in \FGPWa, the  cubic mixing
terms  between $\varphi_{5,6}$ and $\varphi_{j}^2, j =1,2,3$ imply
that if the modes $\varphi_{j}, j =1,2,3$ are excited then
non-normalizable  modes are excited for $\varphi_{j}, j=5,6$.
Indeed, this must happen because of ${\cal N}=1$ supersymmetry:
turning on a fermion means that an entire chiral superfield is becoming
massive, and so the boson masses must be turned on in precisely
the proper manner.

It is not difficult to see that the cubic terms in \asympW\
do this correctly. First, if $\varphi_1$ runs, then only
$\varphi_5$ flows in a non-normalizable manner: and this
is dual to $\cO_1 - \cO_0$ \foot{As explained earlier,
we are always free at add an appropriate amount of the
operator $\cO_0$.}.  This means that the superfield
$\Phi_3$ is developing a mass.  If $\varphi_2$ runs,
then, at lowest order we have ${d \alpha \over d r} \sim
{1 \over 3} \varphi_2^2$ and ${d \beta \over d r} \sim
-\varphi_2^2$, {\it i.e.} to lowest order $\beta =- 3
\alpha$.  From \Lamdefn\ the corresponding $SL(6,\IR)$
scalar is $\Lambda = 2 \alpha \, {\rm diag}
(2 , 2 ,-1, -1,-1,-1)$, which means that the superfield
$\Phi_1$ is developing a mass.  Similarly a flow of
 $\varphi_3$ means that  to lowest order $\beta =+ 3
\alpha$, and $\Phi_2$ is developing a mass.

\subsec{Non-trivial critical points}

The non-trivial, supersymmetric critical point discovered in
\KPW\ corresponds to setting:
\eqn\crtpt{ \alpha= - {1\over 6}\, \log (2)\ ; \quad  \beta = 0  \ ;
\quad \varphi_1 =  {1\over 2}\, \log (3)\ ;
\quad \varphi_j =  0 \ , \ j=2,3,4 \ .}
A careful analysis shows that there are six non-trivial critical
points of the superpotential \superpot, but that
they are all equivalent, and related to \crtpt\ via the permutation
symmetry \permgens.  In terms of the physics on the
brane, they correspond to the choice of superfield that
is to be integrated out to obtain the new phase of
the theory \FGPWa.
The absence of other non-trivial critical points in
the superpotential is consistent with the absence of
other non-trivial IR fixed points of $\cN=4$ Yang-Mills
theory.

\subsec{The Coulomb branch of the
Leigh-Strassler point, and other truncations}

There are several restrictions of our superpotential that
are consistent with the equations of motion.
For example, the $\varphi_j$ only enter into the superpotential
via $\cosh(2\varphi_j)$ and hence
${\del \cW \over \del \varphi_j}$ is proportional
to $\sinh(2\varphi_j)$. This means that it is always consistent to
set any subset of the $\varphi_j$ to zero.

More generally, suppose we set $\varphi_2 = \pm \varphi_3$.
Then, for consistency we must have  
${\del \cW \over \del \varphi_2} = {\del \cW \over \del \varphi_3}$.
This can only be satisfied if either $\varphi_2 = \varphi_3=0$, or
if $\beta \equiv 0$.  One can then check that indeed
${\del \cW \over \del \beta} = 0$ when  one has $\beta =0$ and
$\varphi_2 =  \varphi_3$, and so this is indeed a consistent
truncation.  (Conversely, one can easily show
that it is consistent to set $\beta = 0$ if and only if
$\varphi_2 = \pm \varphi_3$.)
The superpontential reduces to:
\eqn\Wpotred{\widetilde \cW ~=~ - {1 \over 4 \,\rho^4} \,
\big( \cosh(2\varphi_1) - 2 \,
\cosh(2\varphi_2) - \cosh(2\varphi_4) \big)~-~  {1 \over 2}\,
\rho^2 \ \, \big( \cosh(2\varphi_1) + \cosh(2\varphi_4) \big)  .}
If one sets $\varphi_4 =0$, then this superpotential reduces to
that considered \NEMP.  If one further sets $\varphi_2 =0$ then one
gets the superpotential of \FGPWa.  Alternatively,
if  one  sets $\varphi_1= \varphi_4 =0$, one gets the
superpotential considered in \FGPWa.  (Note that the parameter
we call $\alpha$ here is the negative of that used in
\refs{\FGPWa,\NEMP}.)

Another interesting class of truncations are those probing
the Coulomb branch around the massive $\cN \seql 1$ flow
considered in \FGPWa.  That is, set all the
$\varphi_j$ to zero, except  $\varphi_1$.  One then has
the superpotential:
\eqn\Wredtwo{\widehat \cW ~=~  {1 \over 4 \, \rho^4} \,
\big( \cosh(2\varphi_1) - 3 \big)   ~-~  {1 \over 4}\,\rho^2 \,
(\nu^2 + \nu^{-2} ) \, \big( \cosh(2\varphi_1) +1 \big) \,  .}

Setting $\varphi_2 =\varphi_3 =0$ in \asympW, we see that
$\varphi_6$ no longer mixes linearly with other fields, and
so now $\varphi_6$ corresponds to an $AdS_5$ normalizable
mode.  As explained in \FGPWb, $\varphi_6$ represents a
modulus of the Coulomb branch of the $\cN \seql 4$ theory.
In the neighbourhood
of the non-trivial critical point the scaling dimension
of $\varphi_6$ is also $2$, and so from the tables in \FGPWa\
we can identify $\varphi_6$ as being dual to one of the
operators $Tr(\bar \Phi T^A \Phi)$, where $T^A$ are
generators of an $SU(2)$ subgroup of the $SO(6)$ R-symmetry.
Since $\varphi_6 =\sqrt{2}\, \beta$ commutes with $U(1)^3
\subset SO(6)$, it follows that, at the non-trivial
critical point, $\varphi_6$ must be dual
to $Tr(\bar \Phi T^3 \Phi) \sim \Tr (X_1^1 + X_2^2 - X_3^2 -
X_4^2) \equiv \cO_2$ of \scalops.  In other words
$\varphi_6$ is dual to exactly the same operator at both ends
of the flow, suggesting that $\cO_2$ does not mix with
other operators along the flow.  One can also
verify that, all along the flow, the hessian of $\cW$
has a constant eigenvector corresponding to fluctuations
in the $\varphi_6$-direction, which means that $\varphi_6$
does not mix with other fields in supergravity.   We therefore
conclude that $\varphi_6$ represents a Coulomb
modulus for the complete flow.

We have looked at the flows from the non-trivial critical point. 
As is common, we will normalize the coordinate, $r$,
by fixing  $g=2$.
To lowest order, the flows in $\beta$ 
correspond to vevs in $\Tr(\bar \Phi_1 \Phi_1)$
or $\Tr(\bar \Phi_2 \Phi_2)$.   
Generically for $\alpha > \alpha_0 \equiv
- {1\over 6} \log (2)$ one has flows with asymptotic behaviour:
\eqn\alpposinf{\alpha \sim - \coeff{1}{20}\, \log(
\coeff{5}{3} \, r) \,, \quad \beta \sim \pm 3\,
\alpha \,,\quad  \varphi_1  \sim   6\,\alpha \,, 
\quad A \sim -2\, \alpha
\sim \coeff{1}{10}\, \log (\coeff{5}{3} \, r) \,,}
as $r \to 0$.  We have chosen a constant of integration so as
to put the infra-red singularity at $r=0$.
For $\alpha < \alpha_0$ one has flows with asymptotics:
\eqn\alpmininf{ \varphi_1 \to a \,
r^{3 \over 4} \to 0\,, \quad
\alpha \sim  \coeff{1}{4}\, \log\big(
\coeff{4}{3}\,  r \big)\,,
\quad \beta \to  \beta_0\,, \quad A \sim 
\coeff{1}{4}\, \log (r) \,,}
where $a$ and $\beta_0$ are constants.\foot{There are 
other classes of flows in the region in which 
$\cosh(2\varphi_1) >3$, but these are not 
accesible from the non-trivial critical point.}

Between these two classes of flows there are two natural 
ridge lines that asymptote to lines with $\beta = \pm 3 
\alpha$ as $\alpha \to -\infty$.  To get the flows along
these  ridges requires a rather delicate asymptotic
analysis, and one finds that the value of $\varphi_1$ relaxes
to zero {\it very slowly} along the asymptotes.
Again choosing the constant
of integration to put the infra-red singularity at
$r=0$, we find:
\eqn\Coulasymp{\eqalign{\alpha ~\sim~ & {1 \over 4} \, \log(
\coeff{2}{3}\, r) \ , \quad  \beta ~\sim~ \pm( 3 \alpha
+ \varphi_1^2)\ , \cr
\varphi_1^2 ~\sim~ & {1 \over a - 6\,   \log(r) } \ ,
\qquad A(r) ~\sim~ \log(r) \, .}}
for some constant  of integration $a$.
While this suggests that the flow is returning to the Coulomb 
branch of the $\cN \seql 4$ theory in the infra-red, it is
doing it much more slowly than in \alpmininf.  This may 
indicate that some intrisically different physics is to be 
associated with these ridge-line flows.  Note that
the asymptotic behaviour of $e^{2A}$ is also 
significantly different from \alpmininf:  Indeed,
the space-time metric is now asymptotic to:
\eqn\Coulmetric{ ds^2_{1,4} ~\sim~ - dr^2 ~+~ r^2\, 
\eta_{\mu\nu} dx^\mu dx^\nu \,.}
The scale on the brane goes linearly with $r$.  
Amusingly enough, the asymptotic behaviour of 
$\varphi_1^2$ is somewhat reminiscent
of a running gauge coupling.

It was argued in \SGsing\ that the criterion for separating
the physical from the unphysical flows was to require that
the supergravity potential, $\cP$, remain bounded above
along the flows.  This criterion excludes the flows \alpposinf\
in which $\alpha \to +\infty$, and
suggests that the flows with $\alpha \to -\infty$ 
are physical, including the ridge-line flows.
In \SGsing\ the $\cN=1$ flows with $\beta=0$
were considered, and it was further argued that the flows with
$\alpha\to +\infty$ (and $\varphi_1 \not=0$) were unphysical
because they corresponded to giving $\cO_1+2 \cO_0$ a vev or a mass
of the wrong sign, violating positivity.  
It seems reasonable to assume
that something similar is happening here, and  
that the ridge-line flows correspond to the pure
Coulomb branch of the Leigh-Strassler theory, while
the other flows with $\alpha \to -\infty$ correspond to
following a relevant perturbation away from the 
conformal point.  This implies that the ridge line flow
would be given by setting the initial velocities as follows:
\eqn\initvels{{d\, \alpha \over d \,r}\bigg|_{r \to \infty} ~=~
 \cO(e^{- a  \, r})\,,
\qquad {d\, \beta \over d \,r}\bigg|_{r \to \infty} ~\sim~ b\,
e^{- 2\,r}\,, \qquad {d\, \varphi_1\over d \,r}\bigg|_{r \to  
\infty} ~=~ \cO(e^{-  a \,  r})\,,}
for some constant $b$, and for some constant $a$ that
is larger than the largest scaling dimension in the 
$(\alpha,\varphi_1)$-space at the non-trivial
critcal point.   Unfortunately we do not know the
analytic solution.  However, the expansion of the superpotential
near the critical point has terms $\alpha \beta^2$ and
$\varphi_1 \beta^2$, which means that the velocities of 
$\alpha$ and $\varphi_1$ must actually be of order $ b^2 
e^{- 4r}$.  Numerical analysis is consistent with this:
We find that to get the ridge-line flows the initial velocities 
of $\alpha$ and $\varphi_1$ vanish as the square
of the  initial velocity of $\beta$ in the limit
$r \to \infty$.  

We therefore conclude that the two ridges are the 
Coulomb branch of the Leigh-Strassler theory, with
one ridge representing a non-zero vev for
$\Tr(\bar \Phi_1 \Phi_1)$ and the other a non-zero
vev for $\Tr(\bar \Phi_2 \Phi_2)$.

\subsec{A supersymmetric sum rule}

It is elementary to show that the superpotential \superpot\
satisfies the following two identities:
\eqn\hessbits{\sum_{i=1}^4 \, {\partial^2 \cW \over \partial
\varphi_j^2} ~=~ 4 \, \cW \,, \qquad \sum_{i=5}^6 \,
{\partial^2 \cW \over \partial
\varphi_j^2} ~\equiv~{1 \over 6} \,{\partial^2 \cW \over \partial
\alpha^2} +{1 \over 2} \,{\partial^2 \cW \over \partial
\beta^2}  ~=~ {8 \over 3} \, \cW \,\,.}
Consider a general supersymmetric flow:
\eqn\allflow{{d \varphi_j \over d r} \eql  {g \over 2}\,
{\del W \over \del  \varphi_j } \ , \qquad  j=1,\dots,6 \ ,}
and suppose that $\delta \varphi_j$ is some small deviation
away from this flow.  Then to lowest order:
\eqn\varevol{{d  \over d r} \, \delta \varphi_j \eql   {g \over 2}\,
\sum_{k=1}^6 \, {\del^2 W \over \del  \varphi_j \del \varphi_k } \,
\delta \varphi_k \,.}
Now recall that the metric \RGFmetric\ implies that the scale on
the brane is $\mu \equiv e^{A(r)}$, and so $\mu {d \over d \mu} =
{d \over d A}$.  So we change variables to $A$ using \cosmic\ to
obtain:
\eqn\scaling{\mu {d \over d \mu}  \, \delta \varphi_j ~\equiv~
{d  \over d A} \, \delta \varphi_j \eql  - \bigg[{3 \over 2 \cW}\,
\sum_{k=1}^6 \, {\del^2 W \over \del  \varphi_j \del \varphi_k }
\bigg] \,  \delta \varphi_k \,.}
The matrix in square brackets is real and symmetric, and
is thus diagonalizable.  Its eigenvalues represent the scaling
dimensions of the couplings spanned by $\delta \varphi_j$, and
sum of these scaling dimensions is given by the trace of this
matrix.  It follows from this and from \hessbits\
that sum of the scaling dimensions of the fields spanned
by $\delta \varphi_j$ is simply: ${3 \over 2 } \times
(4 + {8 \over 3}) = 10$, and this is true all along the flow,
and not just at critical points.

One can, of course, verify this sum rule at the two
known critical points.  At the $\cN=4$ point the
scaling dimensions of the $\varphi_i$ are $(1,1,1,3,2,2)$,
and at the non-trivial critical point the $\varphi_i$ are
not the diagonal basis, but when one diagonalizes one
finds eigenvalues: $({3 \over 2}, {3 \over 2}, 2,3, 1 +\sqrt{7},
 1 - \sqrt{7})$.  In both cases the sum is $10$.

\newsec{Final Remarks}

In this paper we have constructed a superpotential, $\cW$,
that describes, in five dimensions, the broadest class 
of $\cN =1$ supersymmetric flows so far obtained  
for relevant perturbations of $\cN=4$ supersymmetric 
Yang-Mills theory.  In particular, our superpotential
contains, as special {\it consistent} truncations,
 all the superpotentials considered previously.
We have shown that there are no physically new 
critical points of the superpotential, $\cW$, and thus,
as one would expect from field theory, there
are no new non-trivial supersymmetric fixed points.
On the other hand, while we have not exhibited them here,
the potential, $\cP$, does appear to have some
new non-trivial (non-supersymmetric) critical points.
The more general superpotential has enabled us 
to obtain a sum rule for anomalous dimensions,
and to probe the Coulomb branch of the non-trivial
$\cN=1$ fixed point theory of \RLMS.

It would, of course be interesting to obtain the full
ten-dimensional versions of these flows, 
but the task may be computationally unfeasible.
It is elementary to use the results of \refs{\KPW,\KPNW} to 
compute the metric and dilaton backgrounds.  The
latter is relatively simple.  The complexity comes
not only from the metric, but also from the unknown
tensor gauge field backgrounds.

More practically, one might hope to obtain a better 
understanding of the ten-dimensional origins of
the sum rule, or an understanding of the geometric
structure of the one parameter flow along the Coulomb branch 
of the  $\cN=1$ fixed point theory considered in
section 4.3.  We are currently working on this.
One interesting, and still somewhat surprising feature,
is that if one computes the general dilaton background
for our superpotential, and then seeks out the 
scalar submanifold for which the dilaton/axion background is 
{\it trivial} then one is led to precisely the three scalars: 
$\alpha, \beta, \varphi_1$ with superpotential \Wredtwo.
Thus, this Coulomb branch flow has trivial dilaton/axion
and it would be particularly interesting to see the
ten-dimensional geometric distinction between the
two classes of flows \alpmininf\ and \Coulasymp.

\bigskip
\leftline{\bf Acknowledgements}

We would like to thank K.~Pilch for helpful conversations. 
NW is grateful to the Aspen Center for Physics for its
hospitality while some of the work on this paper was done.  
This work was supported in part by funds provided by the DOE 
under grant number DE-FG03-84ER-40168.

\listrefs
\vfill
\eject
\end